\documentclass[11pt,english,longbibliography]{article}
\usepackage{url}

\usepackage{draft} 
\usepackage{putex}
\usepackage{comment}

\usepackage{amsthm}

\renewcommand{\vec}[1]{\mathbf{#1}}

\newcommand{\CD}{{\cal D}}

\newcommand{\CF}{{\cal F}}

\newcommand{\CS}{{\cal S}}

\makeatletter
\newcommand*{\rom}[1]{\expandafter\@slowromancap\romannumeral #1@}
\makeatother

\usepackage[latin9]{inputenc}
\usepackage{geometry}
\usepackage{verbatim}
\usepackage{prettyref}
\usepackage[numbers,sort&compress]{natbib}
\usepackage{amsmath}
\usepackage{amssymb}
\usepackage{flexisym}
\usepackage{cases}
\usepackage{ytableau}
\usepackage{youngtab}
\usepackage{graphicx}
\usepackage{caption}
\usepackage{subcaption}
\usepackage{tikz}
\usetikzlibrary{arrows.meta,decorations.pathmorphing,decorations.markings}

 \usepackage{slashed}
\usepackage{esvect}
\usepackage{dsfont}

\usepackage{color}
\definecolor{darkgreen}{rgb}{0,0.5,0}
\definecolor{darkblue}{rgb}{0,0,0.6}
\definecolor{purple}{rgb}{0.4,.2,0.7}
\usepackage[colorlinks=true,citecolor=darkgreen,linkcolor=purple,urlcolor=purple]{hyperref}

\numberwithin{equation}{section}
\numberwithin{figure}{section}
\numberwithin{table}{section}

\def\CD{{\cal D}}

\DeclareFontShape{OT1}{cmr}{mx}{n}{<->cmr10}{}

\begin{document}

\fontseries{mx}\selectfont


\title{\centering The phase of de Sitter higher spin gravity
}

\authors{Simone Giombi \worksat{\PUJ} and Zimo Sun \worksat{\PUJ, \IAS} }

\institution{PUJ}{Joseph Henry Laboratories, Princeton University, Princeton, NJ 08544, USA}

\institution{IAS}{Institute for Advanced Study, Princeton, NJ 08540, USA}

\abstract{The one-loop Euclidean partition function on the sphere is known to exhibit a nontrivial phase for massless fields of spin greater than one. Such a phase appears to be in tension with a state counting interpretation of the partition function and its relation to the de Sitter entropy. It has been recently argued that the phase associated with the gravitational path integral can be cancelled by including the contribution of an observer. In this note, we compute the total phase of Vasiliev higher spin gravity on the sphere by summing over the contributions of all spins. We evaluate the resulting infinite sum using two different regularization schemes, obtaining consistent results. We find that for the non-minimal Vasiliev theory, which includes massless fields of all integer spins, the total phase vanishes in all dimensions. This result suggests that the sphere partition function of these theories may be consistent with a counting interpretation, without explicitly including an observer.}

\date{}

\maketitle

\tableofcontents

\section{Introduction}

Understanding quantum gravity in de Sitter space remains a challenging problem, presenting several technical and conceptual issues. At the semiclassical level, 
the one-loop Euclidean gravitational path integral with a positive cosmological constant has been studied by various authors, see e.g. \cite{Hawking:1979ig, Christensen:1979iy, Ginsparg:1982rs, Polchinski:1988ua, Volkov:2000ih}. Due to an infinite number of negative modes in the conformal factor of the metric perturbation, after suitable rotation of the path-integral contour the resulting one-loop partition function generally has a nontrivial
phase. At the leading saddle point, corresponding to a round sphere $S^{d+1}$, Polchinski found the phase $ i^{d+3}=i^1\times i^{d+2}$ of the one-loop partition function, where $i^1$ arises from the constant scalar mode and $i^{d+2}$ is associated with the non-isometric conformal Killing vectors (CKV) \cite{Polchinski:1988ua}. The presence of a non-trivial phase factor is difficult to reconcile with a conventional statistical interpretation of the sphere partition function $Z_{S^{d+1}}$, which is expected to encode the de Sitter entropy $\CS$ by $\CS=\log Z_{S^{d+1}}$ \cite{Gibbons:1976ue, Gibbons:1977mu}. 

It has been recently argued by Maldacena  \cite{Maldacena:2024spf} that this $d$-dependent phase can be removed by including an observer,  modeled as a heavy particle. More precisely, $d$ out of the $d+2$ CKVs that move the observer's trajectory are cancelled by the observer's partition function, leaving a remaining factor of $-i$, which is then removed by imposing the Hamiltonian constraint $H_{\rm tot} = 0$.\footnote{In Ref \cite{Maldacena:2024spf}, imposing the Hamiltonian constraint yields an extra $-i$ and thus an overall minus sign remains. In the later work \cite{Chen:2025jqm}, it is argued that a slightly different way of implementing the Hamiltonian constraint gives $+i$ instead of $-i$. }
Inspired by \cite{Maldacena:2024spf}, there have been many new developments in computing and understanding the phase in various saddle point geometries, such as the direct product of spheres \cite{Ivo:2025yek, Shi:2025amq, Law:2025yec},
 $\mathbb{C}P^2$ \cite{Anninos:2025ltd}, bubble geometry \cite{Ivo:2025fwe}, magnetically charged black holes \cite{Chen:2025jqm,Ivo:2025xek}, and axion wormholes \cite{IvorTangToAppear}.

The appearance of a phase in the sphere partition function is not unique to gravity. It has been shown that the one-loop sphere partition function of massless and partially massless higher spin fields also involves a nontrivial phase \cite{Anninos:2020hfj, Law:2020cpj}. The explicit formula for the higher spin generalization of Polchinski's phase was calculated in these two papers using different methods. We review the derivation of \cite{Anninos:2020hfj} in section \ref{review} below. For a massless field of spin $s\ge 2$ on $S^{d+1}$, the phase $i^{P_s}$ is given by \eqref{Ps}. In general, $P_s$ is a degree $2d-1$ polynomial of $s$. For example, on $S^4$, $P_s = \frac{1}{3}s(s^2-1)^2$ (for $s=2$, this of course reproduces Polchinski's phase factor). 

A consistent interacting theory of higher spin gauge fields necessarily involves an infinite tower of spin content \cite{Vasiliev:1990en, Vasiliev:1992av, Vasiliev:2003ev}. The non-minimal Vasiliev theory of higher spin gravity contains exactly one copy of each massless spinning field, and the minimal Vasiliev theory  contains only even spin fields. In this note, we study the ``total phase" of Vasiliev higher spin theories on $S^{d+1}$, obtained by summing over the contribution of all spins. The primary technical challenge is evaluating the infinite sum $\sum_s P_s$ in a consistent regularization framework. In section \ref{phasecalculation}, we describe two regularization schemes for this sum: The first scheme is the standard dimensional regularization, while the second one is a ``character regularization" which builds on the results of \cite{Anninos:2020hfj}. We find that both regularization schemes yield a vanishing total phase for the non-minimal higher spin gravity in arbitrary dimensions. This implies that the Eucliden sphere partition function of the non-minimal higher spin gravity theory may be consistent with a statistical interpretation on its own, without adding an observer. We also compute the sum over spins in the case of the minimal theory where the odd spin fields are projected out. We find that both regularization schemes yield the same non-vanishing phase $P_{\rm min} = -\frac{\sqrt{\pi}}{2^d\Gamma(\frac{d+1}{2})\Gamma(2-\frac{d}{2})}$ for the minimal higher spin theory. We observe that $P_{\rm min} $ happens to vanish for all odd dimensional spheres except $S^3$. Along the way, we also compute the regularized dimension of the higher spin algebra, which is related to the sum over $s$ of the number of zero modes in each spin-$s$ sector, and controls the dependence of the partition function on Newton's constant.\footnote{This dimension also appears in the AdS free energy of Vasiliev higher spin theory with alternate quantization for all higher spin fields \cite{Giombi:2013yva}, corresponding to gauging the higher spin symmetry in the boundary vector model. It controls the term proportional to $\log(N)$ in the free energy.} We find that in both minimal and non-minimal cases, this regularized sum equals minus the total phase $\sum_s P_s$.  In particular, it vanishes in the non-minimal theory, implying that in this case both phase and $G_N$ dependence disappear. The non-vanishing result we find in the minimal higher spin theory points to an important role played by the odd spin sector in the cancellation mechanism, and it remains to be better understood. 

Higher spin gravity also provides a natural setup to explore the dS/CFT correspondence \cite{Strominger:2001pn}. As proposed in \cite{Anninos:2011ui}, and further refined in \cite{Anninos:2017eib},  the non-minimal higher spin theories containing all integer spins are conjectured to be dual to complex vector models with anticommuting scalars restricted to the $U(N)$ singlet sector, while the minimal theories involve real anticommuting scalars with a $Sp(N)$ singlet constraint. It would be interesting to understand our results for the total phase and $G_N$ dependence on the higher spin side in light of this correspondence. In the case of the non-minimal theory, our findings are potentially consistent with the duality, while the meaning of the result in the minimal theory seems unclear from the point of view of the conjectural dual $Sp(N)$ model. A natural speculation is that the non-vanishing phase might be related to the non-compact nature of the $Sp(N)$ group, which could present additional difficulties in defining such theories, but we leave further investigations of these questions to future work.

\section{Sphere partition function of higher spin gravity}\label{review}

\subsection{Tree-level contribution}
The semiclassical entropy $\CS$ of de Sitter spacetime is given by the Euclidean path integral $\CS = \log Z = \log \int  e^{-I_E[g, \cdots]}$ of some effective field theory $I_E$ of gravity, expanded around the sphere saddle \cite{Gibbons:1976ue, Gibbons:1977mu}. In particular, at tree level, the entropy is minus the on-shell action, i.e., $\CS_0 = - I_E[g_\star, \cdots]$. For Einstein gravity, it gives $\CS_0 = \frac{A}{4G_N}$, where $A$ denotes the area of the static patch horizon.

For Vasiliev higher spin gravity, we cannot compute $\CS_0$ directly since we currently do not know an action. Instead, we can infer the on-shell action by assuming the AdS higher spin holography  \cite{Klebanov:2002ja}, and using the analytical continuation from AdS to dS. More precisely, consider the type-A non-minimal higher spin gravity in AdS$_{d+1}$, that is dual to the free vector $U(N)$ model on the boundary. The higher spin holography predicts $F_{\rm bulk}= F_{U(N)} = 2N F_{\rm conf}$, where $F_{\rm conf}$ denotes the free energy of a conformally coupled scalar on $S^d$. The bulk free energy can be expanded in $G_N$ , namely $F_{\rm bulk} = \frac{1}{G_N} f_0 + f_1 +G_N f_2 +\cdots$, which is also equivalent to loop expansions. For example, $\frac{1}{G_N} f_0 $ is equal to the on-shell action $I_{AdS}$, and $f_1$ is the one-loop free energy, which vanishes \cite{Giombi:2013fka,Giombi:2014iua,Giombi:2014yra, Giombi:2016pvg,Skvortsov:2017ldz,Basile:2018zoy,Basile:2018acb,Sun:2020ame}.
 With the identification of $G_N\sim N^{-1}$, we have $I_{AdS} = 2N F_{\rm conf}$.

Because Vasiliev's master equations take exactly the same form in AdS and dS, up to different choices of reality forms of the master fields, we expect the on-shell action $I_E$ around the sphere saddle to be related to the corresponding AdS action as follows
\begin{align}
I_E = - \frac{\text{vol}(S^{d+1})}{\text{vol}(AdS_{d+1})}I_{AdS} = 2\sin\left(\frac{d\pi}{2}\right) I_{AdS}~. 
\end{align}
It is straightforward to check that such a relation holds in the case of Einstein gravity with a cosmological constant, but for the reasons explained above we expect it to hold in the same way for the higher spin theories (assuming the existence of an underlying action). With the on-shell action known, we can compute the tree-level dS entropy 
\begin{align}\label{CS0}
\text{Non-minimal}:\quad \CS_0 = - I_E =2\times (2N \tilde F_{\rm conf})~,
\end{align}
where $ \tilde F_{\rm conf} \equiv - \sin\left(\frac{d\pi}{2}\right) F_{\rm conf}$. For an arbitrary CFT$_d$, the $\tilde F$ function defined as the sphere free energy multiplied by $- \sin\left(\frac{d\pi}{2}\right)$, is a smooth interpolation between the $a$-anomaly in even $d$ and the sphere free energy in odd $d$ \cite{Giombi:2014xxa}. In particular, for a conformally coupled scalar, it reads \cite{Giombi:2014xxa}.
\begin{align}
\tilde F_{\rm conf} = \frac{1}{\Gamma(d+1)}\int_0^1 du u\sin(\pi u)\Gamma\left(\frac{d}{2}+u\right)\Gamma\left(\frac{d}{2}-u\right)~,
\end{align}
which is manifestly positive for $d\ge 2$. For example, in $3d$, $\tilde F^{3d}_{\rm conf} = \frac{\log (2)}{8}-\frac{3 \zeta (3)}{16 \pi ^2}\approx 0.0638$. The relation \eqref{CS0}, including the factor of 2, is the dimensional continuation of the dS$_4$/CFT$_3$ relation $\CS_{\rm dS} = \log|\Psi_{\rm HH}|^2= 2\log Z_{\rm CFT}(S^3)$ \cite{Maldacena:2011mk, Anninos:2012qw}.

For the minimal Vasiliev theory, the AdS one-loop free energy $f_1$ does not vanish but equals $F_{\rm conf}$. One can attribute it to the renormalization of $G_N$, i.e. $G_N^{-1}\sim N-1$ \cite{Giombi:2013fka}. With this interpretation, the AdS action is $I_{AdS} = (N-1)F_{\rm conf}$. Following the analytical continuation described above, we find the corresponding tree-level dS entropy 
\begin{align}\label{CS0'}
\text{Minimal}:\quad \CS_0 = 2\times (N-1) \tilde F_{\rm conf}~.
\end{align}

 Beyond the on-shell action,  an all-loop  generalization of \eqref{CS0} and \eqref{CS0'} was proposed in \cite{Anninos:2025mje, Dioetal}, and is further tested in supersymmetric Vasiliev theory in  \cite{Dioetal}.

\subsection{One-loop contribution}

A universal formula of one-loop sphere partition function was derived for all parity-invariant effective field theories, in arbitrary dimensions, with arbitrary matter and general gauge symmetry \cite{Anninos:2020hfj}. In particular, for Vasiliev higher spin gravity on $S^{d+1}$, the formula reads
\begin{align}\label{char}
Z^{(1)} = i^{ P}\frac{(2\pi \gamma)^{{\rm dim} G_{\rm HS}}}{{\rm vol}(G_{\rm HS})} {\rm exp}\left[\int_0^\infty \frac{dt}{2t}\frac{1+q}{1-q}\chi_{\rm tot}(t)+\CS_{\rm ct}\right], \quad q\equiv e^{-t}~.
\end{align}
where the UV regulator is suppressed and $\CS_{\rm ct}$ stands for local counterterms. Below, we explain all the ingredients in this formula.
\begin{itemize}
\item $P=\sum_s P_s$ is the total phase, generalizing Polchinski's calculation for Einstein gravity \cite{Polchinski:1988ua}. The sum is for all spin $s\ge 2$ fields in the field spectrum.
The explicit expression of $P_s$ is 
\begin{align}\label{Ps}
P_s = D^{d+3}_{s-1,s-1}+D^{d+3}_{s-2,s-2}- D^{d+2}_{s-1,s-1}~,
\end{align}
where $D^d_{n,s}$ denotes the dimension of the irreducible SO$(d)$ representation with highest weight $(n,s, 0, \cdots,0)$. See appendix \ref{SOd} for more details regarding the dimensions.
We will review the derivation of $P_s$ at the end of this section.
\item $G_{\rm HS}$ denotes the infinite dimensional higher spin group. Its generators correspond to the higher spin Killing tensors on $S^{d+1}$, carrying SO$(d+2) $ representations such as {\tiny\Yvcentermath1 $\yng(1,1)$, $\yng(2,2), \yng(3,3)\cdots$ } It is also the symmetry of the Laplacian on $\mathbb R^d$ \cite{Eastwood:2002su}.
The dimension of $G_{\rm HS}$ is thus ${\rm dim} G_{\rm HS} = \sum_{s\ge 1} D^{d+2}_{s-1,s-1}$, summing over spin $s\ge 1$ fields in the spectrum. In AdS, the  division of group volume is absent because the zero modes are non-normalizable in the standard quantization. Gauging the spin $s$ conserved currect on the boundary of AdS yields $1/\sqrt{N}^{D^{d+2}_{s-1,s-1}}$ in the partition function \cite{Giombi:2013yva} \footnote{After gauging all conserved currents, the coefficient of $\log(N)$ in the AdS free energy becomes $\frac{1}{2}\sum_{s\ge 1} D^{d+2}_{s\!-\!1,s\!-\!1}$. In \cite{Giombi:2013yva}, this sum was computed by using $\zeta$ function regularization. For example, when $d=4$, $\frac{1}{2}D^6_{s\!-\!1,s\!-\!1}=\frac{s^5}{12}+\frac{5 s^4}{24}+\frac{s^3}{6}+\frac{s^2}{24}$ and the $\zeta$ function calculation gives $\frac{\zeta(-5)}{12}+\frac{5 \zeta(-4)}{24}+\frac{\zeta(-3)}{6}+\frac{\zeta(-2)}{24}=\frac{1}{945}$. In section \ref{phasecalculation}, we will show that the sum actually vanishes in dimensional regularization.}. It is in agreement with \eqref{char} provided the identification $G_N\sim 1/N$.
\item $\gamma = \sqrt{\frac{2\pi}{\CS_0}}$, where $\CS_0$ is the on-shell tree-level dS entropy discussed above. It implies the $G_N$ dependence $G_N^{\frac{1}{2}{\rm dim} (G_{\rm HS}) }$ of the one-loop partition function.
\item $\chi_{\rm tot}$ is completely determined by the physical particle spectrum of the theory in dS$_{d+1}$. For example, a conformally coupled scalar on $S^{4}$ has $\chi = \frac{q+q^2}{(1-q)^3}$, which can be reconstructed from a weighted sum over the  quasinormal mode spectrum of the field \cite{Sun:2020sgn}. The non-minimal Vasiliev theory on $S^4$ has $\chi_{\rm tot} = - \frac{q}{(1-q)^2}$.
\end{itemize}

To explain the derivation of $P_s$ in  \cite{Anninos:2020hfj}, we first consider a real massive scalar $\phi$ of mass $m=\sqrt{\frac{d^2}{4}+\nu^2}$ on $S^{d+1}$. 
Starting from the standard heat kernel regularization, the authors \cite{Anninos:2020hfj} found the following expression of the one-loop partition function of $\phi$
\begin{align}\label{massivescalar}
\text{Massive scalar}: \quad \log Z = \int_0^\infty \frac{dt}{2t} \sum_{n\ge 0} D^{d+2}_n \left(q^{\frac{d}{2}+i\nu+n} + q^{\frac{d}{2}-i\nu+n}\right), \quad q= e^{-t}~.
\end{align}
The non-negative integer $n$ labels the spherical harmonics on $S^{d+1}$, which are eigenmodes of the scalar Laplacian. $D^{d+2}_n$ is the multiplicity of spherical harmonics  for a fixed $n$, e.g. $D^3_n=2n+1$. The sum over $n$ can be carried out using \eqref{Dsum}. The higher spin generalization of \eqref{massivescalar} takes a very similar form. More explicitly, the one-loop partition function of a massive spin $s\ge 1$ field on $S^{d+1}$ of mass $m = \sqrt{(d/2+s-2)^2+\nu^2}$ is \cite{Anninos:2020hfj}
\begin{align}\label{massivespin}
\text{Massive spinning}: \quad \log Z = \int_0^\infty \frac{dt}{2t} \sum_{n\ge -1} D^{d+2}_{n,s} \left(q^{\frac{d}{2}+i\nu+n} + q^{\frac{d}{2}-i\nu+n}\right)~,
\end{align}
where $n\ge s$ labels the transverse spin $s$ harmonics on $S^{d+1}$, which carry the $\vec s= (n, s, 0,\cdots, 0)$ representation of SO$(d+2)$. The extension of $n$ all the way to $-1$ captures the contribution of the non-transverse-and-traceless modes, guaranteeing the locality of the final result. 

The massless point of the spin $s$ field is at  $i\nu_s = \frac{d}{2}+s-2 $. Naively plugging $\nu_s$ into \eqref{massivespin} and subtracting the spin $s\!-\!1$ ghost field yields $\log Z \stackrel{\rm naive}{=} \int\frac{dt}{2t}\hat F_s(e^{-t})$, where 
 \begin{align}\label{hF}
\hat F_s(q)= \sum_{n\ge -1} D^{d+2}_{n,s} \left(q^{d+s-2+n} + q^{2-s+n}\right)- \sum_{n\ge -1} D^{d+2}_{n,s-1} \left(q^{d+s-1+n} + q^{1-s+n}\right)~.
\end{align}
For any $s\ge 1$, $\hat F_s(q)$ contains constant terms, causing logarithmic divergence of the integral at large $t$ \footnote{The UV divergence is at $t=0$.}.
These terms correspond to the zero modes of the path integral. For example, at $s=1$ the zero mode is the constant $U(1)$ transformation, and at $s=2$ the zero modes are the Killing vectors on $S^{d+1}$. The correct treatment is dropping the terms proportional to $q^0$ in $\hat F_s(q)$ and dividing the path integral by the gauge group volume.
When $s\ge 2$, $\hat F_s(q)$ also contains negative powers of $q$, leading to exponential divergence of the $t$ integral. As explained in \cite{Anninos:2020hfj}, the negative powers of $q$ (regardless of the sign of the coefficients) are in one-to-one correspondence with the negative modes in the gravitational path integral after Wick rotating the contour.
For instance, when $s=2$, the negative powers of $q$ in $\hat F_2$ is $D^{d+2}_{-1,2}q^{-1} - D^{d+2}_{-1,1}q^{-2}-D^{d+2}_{0,1}q^{-1} = -(d+2)q^{-1} + q^{-2}$, and hence $P_s = d+3$. On the other hand, in the quadratic action of Einstein gravity supplemented by the de Donder gauge fixing term, the traceless and pure-trace parts of the metric fluctuation decouple, with the pure-trace part exhibiting a wrong-sign Gaussian $\int h(\nabla^2 + 2d) h$ \cite{Polchinski:1988ua}. After rotating the contour of the $h$, the modes with angular momentum  $\ell=0, 1$  become negative modes, with the total number being $1+D^{d+2}_1 = d+3$. For an arbitrary $s$, we collect the negative powers of $q$ into 
\begin{align}\label{hFm}
\hat F_s^-(q)&\equiv -\sum_{n=-1}^{s-3} D^{d+2}_{n,s} q^{2-s+n} - \sum_{n=-1}^{s-2} D^{d+2}_{n,s-1} q^{1-s+n}=\sum_{n=0}^{s-2} \left(D^{d+2}_{s-1, n} q^{1-s+n}+D^{d+2}_{s-2,n} q^{n-s}\right)
\end{align}
where we have used $D^{d+2}_{n,s} = -D^{d+2}_{s-1,n+1}$ and $D^{d+2}_{s,s+1} = 0 $. Evaluating $\hat F_s^-(q)$ at $q=1$ and using the branching rule $D^{d+3}_{s,s} = \sum_{n=0}^s D^{d+2}_{s, n}$, we recover $P_s$ in \eqref{Ps}. This number has also been verified by a direct calculation of the one-loop path integral of the Fronsdal action, carried out along the lines of Polchinski \cite{Law:2020cpj}. Roughly speaking, the trace part of the double-traceless Fronsdal field has a wrong-sign action.

Following the prescription of \cite{Ivo:2025yek}, each negative mode contributes a factor of $i$ and hence the phase associated with the massless spin $s$ field is $i^{P_s}$ with $P_s$ given by \eqref{Ps}. In this note, we  focus on computing the regularized sums $P= \sum_s P_s$ and ${\rm dim} G_{\rm HS} = \sum_s D^{d+2}_{s-1,s-1}$, both of which are insensitive to real local finite counterterms.
One might be tempted to use zeta function regularization. It amounts to computing $\sum_s P_s e^{-\delta (s+a)}$ for some $\delta>0$ and $a$, and discarding the divergent terms in the final result in the $\delta\to 0 $ limit. But the result is sensitive to the choice of $a$, and there is no clear justification for any particular choice.

We will therefore adopt two alternative regularization schemes. The first is standard dimensional regularization: we compute an infinite sum in a region of $d$ where the sum is convergent, and then find its analytical continuation at the physical dimensions.  The second, which we refer to as character regularization,
is inspired by the character method introduced in \cite{Anninos:2020hfj} and allows us to work directly in physical dimensions. We will show that these two regularization schemes give the same result.

\section{The total phase}\label{phasecalculation}

In this section, we evaluate the total phase for both non-minimal and minimal Vasiliev theories using dimensional regularization and character regularization. Within the framework of dimensional regularization, we also show that the total phase is related to the dimension of the higher spin algebra.

\subsection{Dimensional regularization}\label{dr}

We first consider the non-minimal Vasiliev theory, which involves all $s\ge 1$ massless spinning fields. In this theory, computing the total phase $\sum_{s\ge 2}P_s$ in dimensional regulatiation is equivalent to evaluating $\sum_{s\ge 0}D^{d+2}_{s,s}$. Using \eqref{Dnsexplicit}, we find the large $s$-asymptotic behavior $D^{d+2}_{s,s}\sim s^{2d-3}$. It implies that the sum $\sum_{s\ge 0}D^{d+2}_{s,s}$ is convergent when ${\rm Re}(d)<1$. In particular, we restrict $d$ to the infinite strip $\CD: -1<{\rm Re}(d)<0$, where the sum over $s$ defines an analytic function. For $d\in \CD$ and $s \ge 2$, the dimension $D^{d+2}_{s,s}$ admits a simple integral representation, c.f. \eqref{Dssint}
\begin{align}\label{Dssint1}
D^{d+2}_{s,s} =\frac{\sin(d\pi)}{\pi}\int_0^\infty \frac{dx}{x^{d+1}}\frac{x+2}{x+1}\left(\frac{D^{d+2}_{s-1}}{(1+x)^{s-1}}\frac{x^2}{(1+x)^2}-\frac{D^{d}_{s}}{(1+x)^s}\right)~,
\end{align}
which allows us to compute the sum over $s\ge 2$ \footnote{The $x$ integral is not well-defined for $s=0, 1$ because the integrand decays as $1/x^{d+s}$ at large $x$.  For the same reason, $s=0$ is excluded from \eqref{tint2}.} before evaluating the $x$ integral 
\begin{align}
\sum_{s\ge 2}D^{d+2}_{s,s} =\frac{\sin(d\pi)}{\pi}\int_0^\infty \frac{dx}{x} \frac{(x+2)((d+2)x+d+1)}{x^{d}(x+1)^3}~,
\end{align}
where we have used \eqref{Dsum}. Since ${\rm Re}(d)<0$, the resulting $x$ integral is convergent and yields $\sum_{s\ge 2}D^{d+2}_{s,s}  = - \frac{d(d+3)}{2}-2$. On the other hand, $D^{d+2}_{0,0} =1$ (corresponding to the trivial representation) and $D^{d+2}_{1,1} = \frac{(d+2)(d+1)}{2}$ (corresponding to the adjoint representation) hold for any $d$. Altogether, $\sum_{s\ge 0} D^{d+2}_{s, s}$ vanishes identically in the strip $\CD$, and thus its analytical continuation vanishes in the whole complex plane of $d$:
\begin{align}\label{main1}
\text{Dim reg}: \quad \sum_{s\ge 0} D^{d+2}_{s, s} =0, \quad d\in\mathbb C~. 
\end{align}
An immediate consequence of \eqref{main1} is the regularized dimension of the higher spin algebra in the non-minimal Vasiliev theory
\begin{align}
 {\rm dim} \left(G^{\text{n-min}}_{\rm HS} \right)= \sum_{s\ge 0} D^{d+2}_{s-1,s-1} = 0~.
\end{align}
So, the one-loop partition function is independent of $G_N$. Similarly, we can show that the total (regularized) phase also vanishes using  \eqref{main1}
\begin{align}
 P_{\text{n-min}}=\sum_{s\ge 2}P_s=\left( 2\sum_{s\ge 0}D^{d+3}_{s,s}-1\right) - \left( \sum_{s\ge 0}D^{d+2}_{s,s}-1\right)=0~.
\end{align}
The absence of the total phase suggests that the dS entropy of  non-minimal Vasiliev theory may be consistent with a counting interpretation without including an observer.

For the minimal Vasiliev theory, which contains only even spin fields, the calculation of the total phase and total dimension of the higher spin algebra is more involved technically. We leave the detailed derivation to the appendix \ref{min} and just present the final result here:
\begin{align}\label{main2}
P_{\text{min}} = -{\rm dim}\left(G^{\text{min}}_{\rm HS}\right)=-\frac{\sqrt{\pi}}{2^d\Gamma(\frac{d+1}{2})\Gamma(2-\frac{d}{2})},
\end{align}
The total phase appears to vanish for even  $d\ge 4$ (odd spacetime dimensions) because of $\Gamma(2-d/2)$ in the denominator. On $S^3$, $P_{\text{min}} = -\frac{1}{2}$.
When $d$ is an odd integer, the total phase is a nontrivial number, but the relation $P_{\rm min} =-{\rm dim}\left(G^{\rm min}_{\rm HS}\right)$ continues to hold.
 We list the value of $P_{\rm min} $ for some small $d$ in the following table
\begin{center}
\begin{tabular}{ |c|c |c| c| c|c|c|}
\hline
 $d$ & 2 & 3 & 5 & 7 &9 &11 \\ 
 \hline
$P_{\rm min} $ & $-\frac{1}{2}$& $-\frac{1}{8}$ & $\frac{1}{128}$& $-\frac{1}{1024}$ & $\frac{5}{32768}$& $-\frac{7}{262144}$  \\
 \hline
\end{tabular}
\end{center}

The vanishing of the total phase and ${\rm dim}(G_{\rm HS})$ that we found is reminiscent of similar cancellations of one-loop free energies, $a$-anomaly coefficients and Casimir energies in the AdS counterpart of the non-minimal higher spin theory \cite{Giombi:2013fka,Giombi:2014iua,Giombi:2014yra, Giombi:2016pvg,Skvortsov:2017ldz,Basile:2018zoy,Basile:2018acb,Sun:2020ame}. In that context, one also finds non-vanishing answers in the minimal theories, but the resulting numbers have a simple boundary interpretation and can be explained by a shift $N\rightarrow N-1$ in the map between $G_N$ and $N$. In the present de Sitter case, the meaning of the numbers we found for $P_{\rm min}$ remains to be understood.

\subsection{Character regularization}

The derivation of $P_s$ reviewed in section \ref{review} naturally leads to a convenient regularization scheme for the sum $\sum_s P_s$, and it gives exactly the same result as the dimensional regularization for $d=3, 4, 5,\cdots$. We illustrate the strategy in the $d=3$ case, which can be easily generalized to higher dimensions.

$P_s$ is encoded in the function  $\hat F^-_s(q)$, c.f. \eqref{hFm}. 
Performing  the sum over $n$ when $d=3$ gives
 \begin{align}
d=3: \quad \hat F_s^-(q) = \frac{\phi_s(q)+q^{-s}(q+1)\varphi_s(q)}{(q-1)^4}, \quad \varphi_s(q) = D^5_{s-2}-D^5_{s-3}q-D^5_{s}q^2+D^5_{s-1}q^3
\end{align}
where $\phi_s(q)$ is a degree 3 polynomial of $q$, whose explicit form is not important. We further define a function by discarding $\phi_s(q)$ in $\hat F_s^-(q) $, i.e., $\CF_s(q)\equiv  \frac{(q+1)\varphi_s(q)}{q^s(q-1)^4}$. As  the degree of $\phi_s$ is 3, $\CF_s$ and $\hat F_s^-(q) $ only differ by terms that are singular at $q=1$. 
Therefore, the constant term of the Laurent expansion of $\CF_s$ around $q=1$ is equal to $\hat F_s^-(1) =P_s $.
The advantage of using  $\CF_s$ over $\hat F_s^-(q) $ is that the former provides a natural way to regularize the sum of $P_s$ because it decays exponentially with $s$ when $q>1$. Evaluating $\sum_{s\ge 2}\CF_s$ using \eqref{Dsum} gives $5\frac{q(q+1)}{(q-1)^4}$, which does not have a constant term in its Laurent expansion around $q=1$.
Thus, we claim that the total phase vanishes for the non-minimal Vasiliev theory on $S^4$, in agreement with dimensional regularization.
Similarly, for the minimal Visiliev theory,  we find 
\small
\begin{align}
\sum_{s\in2\mathbb Z_+} \CF_s (q)&=\frac{q \left(5 q^3+q^2+3 q-1\right)}{(q-1)^4 (q+1)^2}=\frac{1}{2 (q-1)}+\frac{7}{2 (q-1)^2}+\frac{5}{(q-1)^3}+\frac{2}{(q-1)^4}-\frac{1}{8}+\cdots~.
\end{align}
\normalsize
The constant term $-\frac{1}{8}$ again agrees with  dimensional regularization.

The generalization of $\CF_s$ in any  $d\ge 2$ is \footnote{The derivation is slightly different in $d=2$ becomes $\hat F_s$ takes a modified form \cite{Anninos:2020hfj}.}
\begin{align}
\CF_s(q) = \frac{q+1}{q^s(1-q)^{d+1}}\left( D^{d+2}_{s-2}-D^{d+2}_{s-3}q-D^{d+2}_{s}q^2+D^{d+2}_{s-1}q^3\right)~.
\end{align}
It is straightforward to check that this function yields the same phase as dimensional regularization for both minimal and non-minimal Vasiliev theories in  $d\ge 2$.

Let's also mention that the simpler regularized sum $\sum_s P_s \, q^s$ with $|q|<1$ gives the same result for the total phase. More explicitly, at $d=3$, we have 
\begin{align}
&\sum_{s\ge 2} P_s \, q^s = \frac{40}{(q-1)^6}+\frac{120}{(q-1)^5}+\frac{126}{(q-1)^4}+\frac{52}{(q-1)^3}+\frac{6}{(q-1)^2}+O\left(q-1\right)\nonumber\\
&\sum_{s\in 2\mathbb Z_+} P_s \, q^s = \frac{20}{(q-1)^6}+\frac{60}{(q-1)^5}+\frac{63}{(q-1)^4}+\frac{26}{(q-1)^3}+\frac{3}{(q-1)^2}-\frac{1}{8}+O\left(q-1\right)~.
\end{align}

\section*{Acknowledgement}
We thank D. Anninos,  I. Klebanov and B. Muehlmann for useful discussions.
Z.S. is supported by the U.S. Department of Energy grant
DE-SC0009988.

\appendix

\section{Some useful formulas for representation dimensions}\label{SOd}
Irreducible representations of SO$(d)$ are labelled by a highest weight vector $\vec s = (s_1, s_2, \cdots)$. Their dimensions $D^d_{\vec s}$ are given by the Weyl dimension formula.
We are mainly interested in the case of $(s, 0, 0,\cdots)$ and $(n,s,0,\cdots)$, where $n\ge s$ are non-negative integers. 
In the former case, corresponding to the spin $s$ representation of SO$(d)$, the dimension is
\begin{align}\label{Dds}
D^d_s =\binom{s+d-1}{d-1}- \binom{s+d-3}{d-1} =(2s+d-2)\frac{\Gamma(s+d-2)}{\Gamma(d-1)\Gamma(s+1)}~.
\end{align}
In representation theory, $s\in\mathbb N$, but \eqref{Dds} can be analytically continued to the full complex plane of $s$ (and also $d$). In particular, it implies $D^d_s$ vanishes when $s$ is a negative integer. 

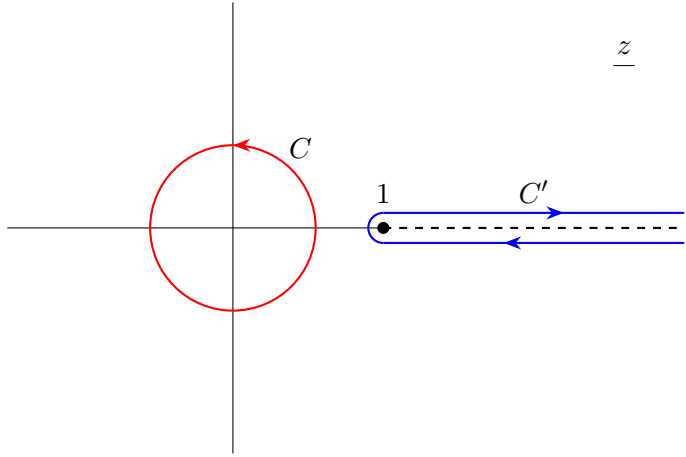
\begin{figure}[t]
\centering
\begin{tikzpicture}[scale=2]

\node at (2.6,1.2) {$z$};
\draw (2.53,1.08) -- (2.67,1.08);
\draw(-1.5,0) -- (1,0);
\draw (0,-1.5) -- (0,1.5);
\draw[red, thick,
      postaction={decorate},
      decoration={markings,
        mark=at position 0.25 with {\arrow{Stealth}}
      }]
      (0,0) circle (0.55);
\draw[black, thick, dashed] (1,0) -- (3,0);
\draw[blue, thick,
      postaction={decorate},
      decoration={markings,
        mark=at position 0.6 with {\arrow{Stealth}}}]
      (1,0.1) -- (3,0.1);

\draw[blue, thick,
      postaction={decorate},
      decoration={markings,
        mark=at position 0.6 with {\arrow{Stealth}}}]
      (3,-0.1) -- (1,-0.1);

\draw[blue, thick] (1,0.1) arc (90:270:0.1);

\fill (1,0) circle (0.04);
\node[above] at (1,0.1) {$1$};

\node[above] at (0.45,0.4) {$C$};
\node[above] at (2,0.1) {$C'$};

\end{tikzpicture}
\caption{Contour deformation of the $z$ integral. The red line $C$ denotes the original contour within the unit disk, and the blue line $C'$ represents the deformed contour along the branch cut.}
\label{Contour}
\end{figure}

Treating $d$ as a continuous variable in $\mathbb C$, we derive an integral representation of $D^d_s$ in the strip $2-s<{\rm Re}(d)<2$ where $s\in\mathbb Z_+$. We first group all $D^d_s$ into a single generating function
\begin{align}\label{Dsum}
\sum_{s\ge 0}D^{d}_s z^s = \frac{1+z}{(1-z)^{d-1}}, \quad 0<|z|<1~.
\end{align} 
$D^d_s$ can thus be computed as a contour integral, namely
$
D^{d}_{s} = \int_{C}\frac{dz}{2\pi i }  \frac{1+z}{z^{s+1}(1-z)^{d-1}}
$, 
where $C$ is a circle within the unit disk centered at the origin. Because  $d$ is a continuous variable, the integrand has a branch cut from $z=1$ to $\infty$, shown as the dashed line in fig. \ref{Contour}. We then deform the circular contour $C$ to a contour $C'$ (the blue line in fig. \ref{Contour}) that runs along the branch cut. Since there are no poles in the region bounded by $C$ and $C'$, the new contour yields
\begin{align}\label{Ddsint}
D_s^d = -\frac{\sin(d\pi)}{\pi}\int_0^\infty \frac{dx}{x^{d-1}}\frac{2+x}{(1+x)^{s+1}}~.
\end{align} 
This integral is convergent provided $2-s<{\rm Re}(d)<2$.

Similarly, we can write down $D^d_{n,s}$ in terms of $\Gamma$ functions using the Weyl dimensional formula
\begin{align}\label{Dnsexplicit}
D^d_{n,s}=\frac{(d+2 n-2) (d+2 s-4) (n-s+1) (d+n+s-3) \Gamma (d+n-3) \Gamma (d+s-4)}{\Gamma (d-3) \Gamma (d-1) \Gamma (n+2) \Gamma
   (s+1)}~.
\end{align}
 Here, we give another useful expression which relates different representations \cite{Anninos:2020hfj}
\begin{align}\label{DDD}
D^{d+2}_{n,s} = D^{d+2}_n D^d_s- D^{d+2}_{s-1}D^d_{n+1}~.
\end{align}
Combining \eqref{Ddsint} and \eqref{DDD} immediately yields an integral representation of $D^{d+2}_{n,s }$
\begin{align}\label{Dssint}
D^{d+2}_{n,s} = \frac{\sin(d\pi)}{\pi}\int_0^\infty \frac{dx}{x}\frac{2+x}{x^{d}(1+x)^{n+1}}\left(D^{d+2}_{s-1}\frac{x^2}{1+x}-D^{d}_{s}\right)~,
\end{align}
which is convergent in the strip $1-n<{\rm Re}(d)<0$. The poles of $D^{d+2}_{s-1}$ and $D^{d}_{s}$ at negative integer $d$ are cancelled by $\sin(d\pi)$.

\section{Dimensional regularization of minimal Vasiliev theory}\label{min}
In the minimal Vasiliev theory, we need to compute the sum of $D^{d+2}_{s,s}$ over $s=2,4,6,\cdots,\infty$.
The same strategy that was used to prove $\sum_{s\ge 0}D^{d+2}_{s,s} =0$ in section \ref{dr} can also be applied here.
For $d\in \CD$, we use the integral representation \eqref{Dssint1} of $D^{d+2}_{s,s}$ and perform the sum inside the integral
\begin{align}\label{tint2}
\sum_{s\in2\mathbb Z_+}D^{d+2}_{s,s} =\frac{\sin(d\pi)}{\pi}\int_0^\infty dx\left(\frac{x+2}{x^{d+1}(1+x)} - \frac{x(x+2)+2}{(x(x+2))^d(x+1)^{3-d}}\right)~.
\end{align}
The sum of the integrand is evaluated using  \eqref{Dsum} and 
\begin{align}
\sum_{s\in 2\mathbb N}D^{d+2}_{s}z^s = \frac{1}{2}\frac{1+z}{(1-z)^{d+1}}+\frac{1}{2}\frac{1-z}{(1+z)^{d+1}}, \quad 0<|z|<1~.
\end{align}
The integral \eqref{tint2} can be computed analytically  after making the substitution $x\to e^{\frac{y}{2}}-1$, and the final result reads $ \sum_{s\in2\mathbb Z_+} D^{d+2}_{s,s}=-1+\Psi(d)$, where 
\begin{align}
\Psi(d)= -\frac{\sqrt{\pi}}{2^d\Gamma(\frac{d+1}{2})\Gamma(2-\frac{d}{2})}~.
\end{align}

The calculation above established the relation $ \sum_{s\in2\mathbb N} D^{d+2}_{s,s}=\Psi(d) $
in the strip $-1<{\rm Re}(d)<0$. On the other hand, the function $\Psi(d)$ obviously admits analytical continuation in  the whole complex plane of $d$. Therefore, in the framework of  dimensional regularization, we have 
\begin{align}\label{sumin}
\text{Dim reg}: \quad \sum_{s\in 2\mathbb N} D^{d+2}_{s, s} =\Psi(d),  \quad d\in\mathbb C~.
\end{align}
Combining \eqref{main1} and \eqref{sumin} leads to the result \eqref{main2}.

\bibliography{refs}

\providecommand{\href}[2]{#2}\begingroup\raggedright\begin{thebibliography}{10}

\bibitem{Hawking:1979ig}
S.~W. Hawking and W.~Israel, {\em {General Relativity}: {An Einstein Centenary
  Survey}}.
\newblock Univ. Pr., Cambridge, UK, 1979.

\bibitem{Christensen:1979iy}
S.~M. Christensen and M.~J. Duff, ``{Quantizing Gravity with a Cosmological
  Constant},'' \href{http://dx.doi.org/10.1016/0550-3213(80)90423-X}{{\em Nucl.
  Phys. B} {\bfseries 170} (1980) 480--506}.

\bibitem{Ginsparg:1982rs}
P.~H. Ginsparg and M.~J. Perry, ``{Semiclassical Perdurance of de Sitter
  Space},'' \href{http://dx.doi.org/10.1016/0550-3213(83)90636-3}{{\em Nucl.
  Phys. B} {\bfseries 222} (1983) 245--268}.

\bibitem{Polchinski:1988ua}
J.~Polchinski, ``{The phase of the sum over spheres},''
  \href{http://dx.doi.org/10.1016/0370-2693(89)90387-0}{{\em Phys. Lett. B}
  {\bfseries 219} (1989) 251--257}.

\bibitem{Volkov:2000ih}
M.~S. Volkov and A.~Wipf, ``{Black hole pair creation in de Sitter space: A
  Complete one loop analysis},''
  \href{http://dx.doi.org/10.1016/S0550-3213(00)00287-X}{{\em Nucl. Phys. B}
  {\bfseries 582} (2000) 313--362},
  \href{http://arxiv.org/abs/hep-th/0003081}{{\ttfamily arXiv:hep-th/0003081}}.

\bibitem{Gibbons:1976ue}
G.~W. Gibbons and S.~W. Hawking, ``{Action Integrals and Partition Functions in
  Quantum Gravity},'' \href{http://dx.doi.org/10.1103/PhysRevD.15.2752}{{\em
  Phys. Rev. D} {\bfseries 15} (1977) 2752--2756}.

\bibitem{Gibbons:1977mu}
G.~W. Gibbons and S.~W. Hawking, ``{Cosmological Event Horizons,
  Thermodynamics, and Particle Creation},''
  \href{http://dx.doi.org/10.1103/PhysRevD.15.2738}{{\em Phys. Rev. D}
  {\bfseries 15} (1977) 2738--2751}.

\bibitem{Maldacena:2024spf}
J.~Maldacena, ``{Real observers solving imaginary problems},''
  \href{http://arxiv.org/abs/2412.14014}{{\ttfamily arXiv:2412.14014
  [hep-th]}}.

\bibitem{Chen:2025jqm}
Y.~Chen, D.~Stanford, H.~Tang, and Z.~Yang, ``{On the phase of the de Sitter
  density of states},'' \href{http://arxiv.org/abs/2511.01400}{{\ttfamily
  arXiv:2511.01400 [hep-th]}}.

\bibitem{Ivo:2025yek}
V.~Ivo, J.~Maldacena, and Z.~Sun, ``{Physical instabilities and the phase of
  the Euclidean path integral},''
  \href{http://arxiv.org/abs/2504.00920}{{\ttfamily arXiv:2504.00920
  [hep-th]}}.

\bibitem{Shi:2025amq}
X.~Shi and G.~J. Turiaci, ``{The phase of the gravitational path integral},''
  \href{http://dx.doi.org/10.1007/JHEP07(2025)047}{{\em JHEP} {\bfseries 07}
  (2025) 047}, \href{http://arxiv.org/abs/2504.00900}{{\ttfamily
  arXiv:2504.00900 [hep-th]}}.

\bibitem{Law:2025yec}
Y.~T.~A. Law and V.~Lochab, ``{Gravitons on Nariai Edges},''
  \href{http://arxiv.org/abs/2506.02142}{{\ttfamily arXiv:2506.02142
  [hep-th]}}.

\bibitem{Anninos:2025ltd}
D.~Anninos, C.~Baracco, S.~Brian, and F.~Denef, ``{Features of the Partition
  Function of a $\Lambda>0$ Universe},''
  \href{http://arxiv.org/abs/2505.11330}{{\ttfamily arXiv:2505.11330
  [hep-th]}}.

\bibitem{Ivo:2025fwe}
V.~Ivo, ``{One loop aspects of Coleman de Luccia instantons at small
  backreaction},'' \href{http://arxiv.org/abs/2509.18651}{{\ttfamily
  arXiv:2509.18651 [hep-th]}}.

\bibitem{Ivo:2025xek}
V.~Ivo and Z.~Sun, ``{The phase of charged Nariai solutions},''
  \href{http://arxiv.org/abs/2511.06604}{{\ttfamily arXiv:2511.06604
  [hep-th]}}.

\bibitem{IvorTangToAppear}
V.~Ivor and H.~Tang, ``{One-loop aspects of de Sitter Axion wormholes},'' 2026.
\newblock to appear.

\bibitem{Anninos:2020hfj}
D.~Anninos, F.~Denef, Y.~T.~A. Law, and Z.~Sun, ``{Quantum de Sitter horizon
  entropy from quasicanonical bulk, edge, sphere and topological string
  partition functions},'' \href{http://dx.doi.org/10.1007/JHEP01(2022)088}{{\em
  JHEP} {\bfseries 01} (2022) 088},
  \href{http://arxiv.org/abs/2009.12464}{{\ttfamily arXiv:2009.12464
  [hep-th]}}.

\bibitem{Law:2020cpj}
Y.~T.~A. Law, ``{A compendium of sphere path integrals},''
  \href{http://dx.doi.org/10.1007/JHEP12(2021)213}{{\em JHEP} {\bfseries 12}
  (2021) 213}, \href{http://arxiv.org/abs/2012.06345}{{\ttfamily
  arXiv:2012.06345 [hep-th]}}.

\bibitem{Vasiliev:1990en}
M.~A. Vasiliev, ``{Consistent equation for interacting gauge fields of all
  spins in (3+1)-dimensions},''
  \href{http://dx.doi.org/10.1016/0370-2693(90)91400-6}{{\em Phys. Lett. B}
  {\bfseries 243} (1990) 378--382}.

\bibitem{Vasiliev:1992av}
M.~A. Vasiliev, ``{More on equations of motion for interacting massless fields
  of all spins in (3+1)-dimensions},''
  \href{http://dx.doi.org/10.1016/0370-2693(92)91457-K}{{\em Phys. Lett. B}
  {\bfseries 285} (1992) 225--234}.

\bibitem{Vasiliev:2003ev}
M.~A. Vasiliev, ``{Nonlinear equations for symmetric massless higher spin
  fields in (A)dS(d)},''
  \href{http://dx.doi.org/10.1016/S0370-2693(03)00872-4}{{\em Phys. Lett. B}
  {\bfseries 567} (2003) 139--151},
  \href{http://arxiv.org/abs/hep-th/0304049}{{\ttfamily arXiv:hep-th/0304049}}.

\bibitem{Giombi:2013yva}
S.~Giombi, I.~R. Klebanov, S.~S. Pufu, B.~R. Safdi, and G.~Tarnopolsky, ``{AdS
  Description of Induced Higher-Spin Gauge Theory},''
  \href{http://dx.doi.org/10.1007/JHEP10(2013)016}{{\em JHEP} {\bfseries 10}
  (2013) 016}, \href{http://arxiv.org/abs/1306.5242}{{\ttfamily arXiv:1306.5242
  [hep-th]}}.

\bibitem{Strominger:2001pn}
A.~Strominger, ``{The dS / CFT correspondence},''
  \href{http://dx.doi.org/10.1088/1126-6708/2001/10/034}{{\em JHEP} {\bfseries
  10} (2001) 034}, \href{http://arxiv.org/abs/hep-th/0106113}{{\ttfamily
  arXiv:hep-th/0106113}}.

\bibitem{Anninos:2011ui}
D.~Anninos, T.~Hartman, and A.~Strominger, ``{Higher Spin Realization of the
  dS/CFT Correspondence},''
  \href{http://dx.doi.org/10.1088/1361-6382/34/1/015009}{{\em Class. Quant.
  Grav.} {\bfseries 34} no.~1, (2017) 015009},
  \href{http://arxiv.org/abs/1108.5735}{{\ttfamily arXiv:1108.5735 [hep-th]}}.

\bibitem{Anninos:2017eib}
D.~Anninos, F.~Denef, R.~Monten, and Z.~Sun, ``{Higher Spin de Sitter Hilbert
  Space},'' \href{http://dx.doi.org/10.1007/JHEP10(2019)071}{{\em JHEP}
  {\bfseries 10} (2019) 071}, \href{http://arxiv.org/abs/1711.10037}{{\ttfamily
  arXiv:1711.10037 [hep-th]}}. [Erratum: JHEP 06, 085 (2024)].

\bibitem{Klebanov:2002ja}
I.~R. Klebanov and A.~M. Polyakov, ``{AdS dual of the critical O(N) vector
  model},'' \href{http://dx.doi.org/10.1016/S0370-2693(02)02980-5}{{\em Phys.
  Lett. B} {\bfseries 550} (2002) 213--219},
  \href{http://arxiv.org/abs/hep-th/0210114}{{\ttfamily arXiv:hep-th/0210114}}.

\bibitem{Giombi:2013fka}
S.~Giombi and I.~R. Klebanov, ``{One Loop Tests of Higher Spin AdS/CFT},''
  \href{http://dx.doi.org/10.1007/JHEP12(2013)068}{{\em JHEP} {\bfseries 12}
  (2013) 068}, \href{http://arxiv.org/abs/1308.2337}{{\ttfamily arXiv:1308.2337
  [hep-th]}}.

\bibitem{Giombi:2014iua}
S.~Giombi, I.~R. Klebanov, and B.~R. Safdi, ``{Higher Spin AdS$_{d+1}$/CFT$_d$
  at One Loop},'' \href{http://dx.doi.org/10.1103/PhysRevD.89.084004}{{\em
  Phys. Rev. D} {\bfseries 89} no.~8, (2014) 084004},
  \href{http://arxiv.org/abs/1401.0825}{{\ttfamily arXiv:1401.0825 [hep-th]}}.

\bibitem{Giombi:2014yra}
S.~Giombi, I.~R. Klebanov, and A.~A. Tseytlin, ``{Partition Functions and
  Casimir Energies in Higher Spin AdS$_{d+1}$/CFT$_d$},''
  \href{http://dx.doi.org/10.1103/PhysRevD.90.024048}{{\em Phys. Rev. D}
  {\bfseries 90} no.~2, (2014) 024048},
  \href{http://arxiv.org/abs/1402.5396}{{\ttfamily arXiv:1402.5396 [hep-th]}}.

\bibitem{Giombi:2016pvg}
S.~Giombi, I.~R. Klebanov, and Z.~M. Tan, ``{The ABC of Higher-Spin AdS/CFT},''
  \href{http://dx.doi.org/10.3390/universe4010018}{{\em Universe} {\bfseries 4}
  no.~1, (2018) 18}, \href{http://arxiv.org/abs/1608.07611}{{\ttfamily
  arXiv:1608.07611 [hep-th]}}.

\bibitem{Skvortsov:2017ldz}
E.~D. Skvortsov and T.~Tran, ``{AdS/CFT in Fractional Dimension and Higher Spin
  Gravity at One Loop},'' \href{http://dx.doi.org/10.3390/universe3030061}{{\em
  Universe} {\bfseries 3} no.~3, (2017) 61},
  \href{http://arxiv.org/abs/1707.00758}{{\ttfamily arXiv:1707.00758
  [hep-th]}}.

\bibitem{Basile:2018zoy}
T.~Basile, E.~Joung, S.~Lal, and W.~Li, ``{Character Integral Representation of
  Zeta function in AdS$_{d+1}$: I. Derivation of the general formula},''
  \href{http://dx.doi.org/10.1007/JHEP10(2018)091}{{\em JHEP} {\bfseries 10}
  (2018) 091}, \href{http://arxiv.org/abs/1805.05646}{{\ttfamily
  arXiv:1805.05646 [hep-th]}}.

\bibitem{Basile:2018acb}
T.~Basile, E.~Joung, S.~Lal, and W.~Li, ``{Character integral representation of
  zeta function in AdS$_{d+1}$. Part II. Application to partially-massless
  higher-spin gravities},''
  \href{http://dx.doi.org/10.1007/JHEP07(2018)132}{{\em JHEP} {\bfseries 07}
  (2018) 132}, \href{http://arxiv.org/abs/1805.10092}{{\ttfamily
  arXiv:1805.10092 [hep-th]}}.

\bibitem{Sun:2020ame}
Z.~Sun, ``{AdS one-loop partition functions from bulk and edge characters},''
  \href{http://dx.doi.org/10.1007/JHEP12(2021)064}{{\em JHEP} {\bfseries 12}
  (2021) 064}, \href{http://arxiv.org/abs/2010.15826}{{\ttfamily
  arXiv:2010.15826 [hep-th]}}.

\bibitem{Giombi:2014xxa}
S.~Giombi and I.~R. Klebanov, ``{Interpolating between $a$ and $F$},''
  \href{http://dx.doi.org/10.1007/JHEP03(2015)117}{{\em JHEP} {\bfseries 03}
  (2015) 117}, \href{http://arxiv.org/abs/1409.1937}{{\ttfamily arXiv:1409.1937
  [hep-th]}}.

\bibitem{Maldacena:2011mk}
J.~Maldacena, ``{Einstein Gravity from Conformal Gravity},''
  \href{http://arxiv.org/abs/1105.5632}{{\ttfamily arXiv:1105.5632 [hep-th]}}.

\bibitem{Anninos:2012qw}
D.~Anninos, ``{De Sitter Musings},''
  \href{http://dx.doi.org/10.1142/S0217751X1230013X}{{\em Int. J. Mod. Phys. A}
  {\bfseries 27} (2012) 1230013},
  \href{http://arxiv.org/abs/1205.3855}{{\ttfamily arXiv:1205.3855 [hep-th]}}.

\bibitem{Anninos:2025mje}
D.~Anninos, C.~Baracco, V.~A. Letsios, and G.~A. Silva, ``{Fermionic fields of
  higher spin in de Sitter space},''
  \href{http://arxiv.org/abs/2510.19652}{{\ttfamily arXiv:2510.19652
  [hep-th]}}.

\bibitem{Dioetal}
D.~Anninos, C.~Baracco, B.~Muehlmann, and V.~A. Letsios, ``{Toward a
  microscopic completion of $S^4$ higher spin gravity},'' 2026.
\newblock to appear.

\bibitem{Eastwood:2002su}
M.~G. Eastwood, ``{Higher symmetries of the Laplacian},''
  \href{http://dx.doi.org/10.4007/annals.2005.161.1645}{{\em Annals Math.}
  {\bfseries 161} (2005) 1645--1665},
  \href{http://arxiv.org/abs/hep-th/0206233}{{\ttfamily arXiv:hep-th/0206233}}.

\bibitem{Sun:2020sgn}
Z.~Sun, ``{Higher spin de Sitter quasinormal modes},''
  \href{http://dx.doi.org/10.1007/JHEP11(2021)025}{{\em JHEP} {\bfseries 11}
  (2021) 025}, \href{http://arxiv.org/abs/2010.09684}{{\ttfamily
  arXiv:2010.09684 [hep-th]}}.

\end{thebibliography}\endgroup
\bibliographystyle{utphys}

\end{document}